\documentclass{aa}
\usepackage{graphicx}

\begin{document}

\title{Photometric study of two novae in M31}

\author{Y. C. Joshi\inst{1}, A. K. Pandey\inst{1}, D. Narasimha\inst{2},
Y. Giraud-H\'{e}raud\inst{3}, R. Sagar\inst{1} and J. Kaplan\inst{3}}

\offprints{Y. C. Joshi, \\
           \email{yogesh@upso.ernet.in}
            }

\institute{ State Observatory, Manora peak, Nainital - 263129, Uttaranchal, 
India 
          \and
              Tata Institute of Fundamental Research, Homi Bhabha Road, 
Mumbai -- 400 005, India
           \and
              Laboratoire de Physique Corpusculaire, College de France, 
Laboratoire associe au CNRS-IN2P3 (URA 6411), 11 place Marcelin Berthelot, 
75231 Paris Cedex 05, France
              }

\date{Received 26 May 2003  /accepted 15 October 2003}

\abstract{
We report $R_cI_c$ light curves of 2 novae in the M31 galaxy which were
detected in the four year Nainital Microlensing Survey. One of these novae
has been tracked from the initial
increase in flux while other has been observed during its descending phase
of brightness. The photometry of the first nova during the outburst phase
suggests its peak $R_c$ magnitude to be about 17.2 mag with a flux decline
rate of 0.11 mag day$^{-1}$ which indicates that it was a fast nova. A month
after its outburst, it shows reddening followed by a plateau in
$I_c$ flux. The second nova exhibits a bump in $R_c$ and $I_c$, possibly
about three weeks after the outburst.
}
\authorrunning{Y. C. Joshi et al.}
\titlerunning{Photometric study of two novae in M31}
\maketitle
\section{Introduction}
Cataclysmic variables (CVs) are close binary systems consisting of a white
dwarf primary and a late-type main sequence secondary star. Novae are a
sub-class of cataclysmic variables characterized by the presence of a sudden
increase of brightness, called outbursts, due to thermonuclear
runway in the envelope of the primary, causing the system brightness to
increase typically by 10-20 mag. These are bright objects which reach up
to $M_{V} \sim$ -9.0 mag at maximum and their rate of decline is tightly
correlated with their absolute magnitude at maximum (McLaughlin 1945). The
study of novae in
external galaxies is important to infer their distances as these objects are
one of the brightest standard candles up to the Virgo cluster (cf. Jacoby et al.
1992 for a review) and tracers of differences in the stellar content among
galaxies (cf. Van den Bergh 1988 for a review).

M31, our nearest large galaxy, has been a target of searches for novae since
the pioneering work of Hubble (1929). Later Arp (1956), Rosino (1964, 1973),
Rosino et al. (1989), Ciardullo et al. (1987), Sharov \& Alksnis (1991),
Tomaney \& Shafter (1992), Rector et al. (1999) and Shafter \& Irby (2001)
have extended the systematic search for novae in M31.
In collaboration with the AGAPE (Andromeda Gravitational Amplification Pixel
Experiment) group, we started Cousins $R$ and $I$ photometric observations
of M31 in 1998 to search for microlensing events. Based on the
4 year observations, we have already reported the discovery of new Cepheids
and other variable stars (Joshi et al. 2003a). As a microlensing survey
program is ideally suited to monitor the flux and temperature variation during 
transient events, we have extended our search to detect nova outbursts.
Here we report photometric light curves of two novae detected in the target
field, one each in 2000 and 2001 observing seasons. We show that an increase
in flux at longer wavelengths a few weeks after the initial rapid
decline is a common phenomenon in the novae light curves.
\section{Observations}
We have undertaken a program called the {\bf Nainital Microlensing Survey} to
detect microlensing events in the direction of M31, at the State Observatory,
Nainital, India, since 1998. Cousins $R$ and $I$ broad band CCD observations
of M31 were carried out for an $\sim6'\times 6'$ field in 1998 and an
$\sim13'\times 13'$ field during 1999 to 2001 observing seasons using the
104-cm Sampurnanand Telescope at the f/13 Cassegrain focus. The total
integrated observing time devoted to the survey ranges from $\sim$ 30 minutes
to 2 hours each night. The $13'\times 13'$ target field ($\alpha _{2000}$ =
$0^{h} 43^{m} 38^{s}$ and $\delta_{2000}$ = $+41^{\circ}09^{\prime}.1$) is
centered at a distance of $\sim$ 15 arcmin away from the center of M31. The
average seeing during the 141 observed nights spanning 4 years was $\sim$ 2
arcsec.  An overview of the observational detail has been given by Joshi et al.
(2003a).
\setcounter{figure}{0}
\begin{figure*}[ht]
\centering
\includegraphics[height=6.0cm,width=6.0cm]{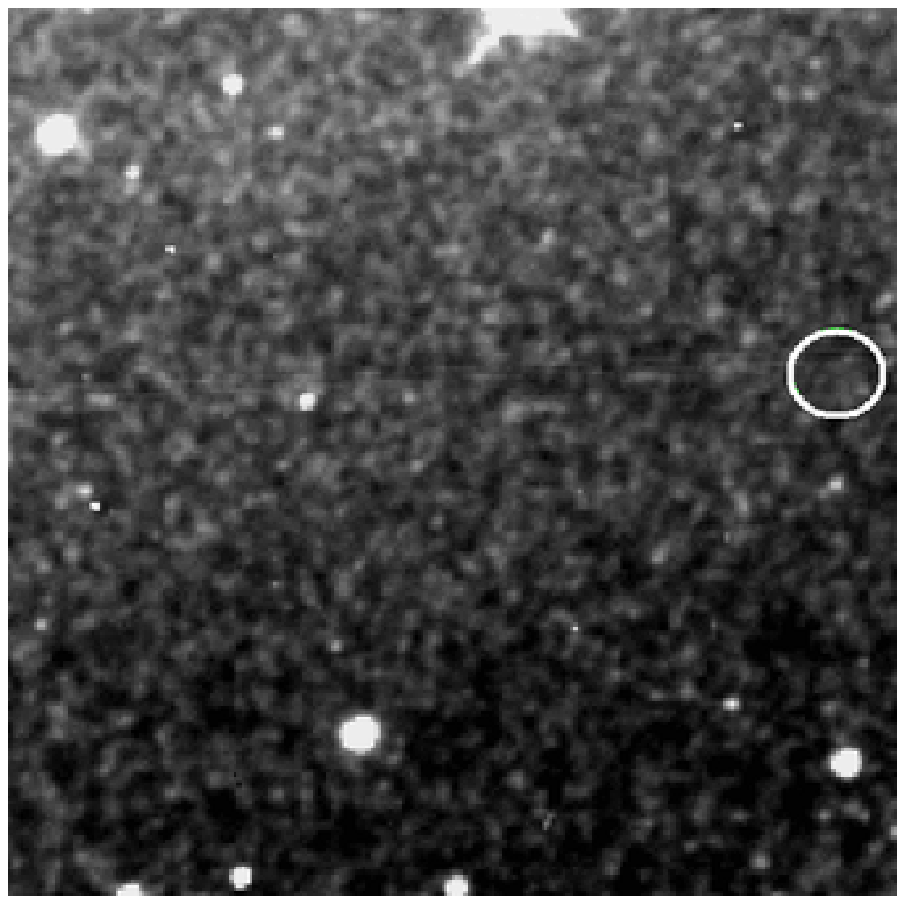}\hspace{1cm} \includegraphics[height=6.0cm,width=6.0cm]{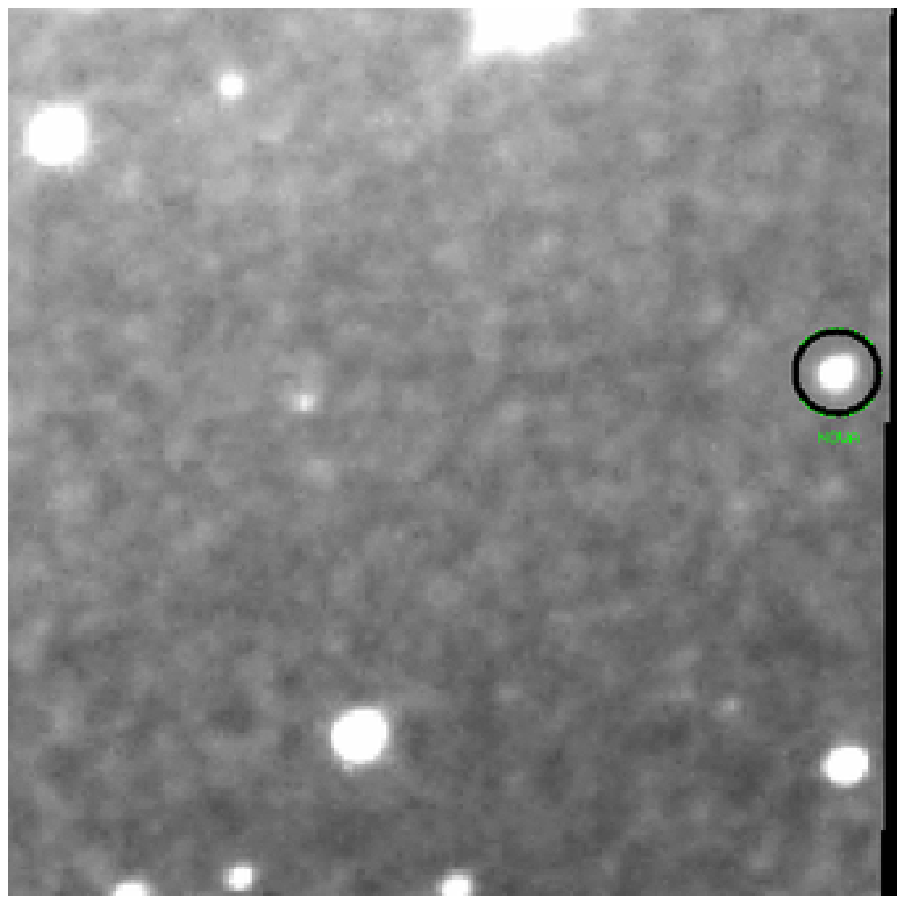}
\caption{A 2 arcmin wide subset of two different $R$ band images taken on
January 11, 2000 (left panel) and October 20, 2000 (right panel). North is at 
the top and East is to the left. The left frame shows no star at the position
marked by a circle while right frame shows a star (nova NMS-1) of brightness
$R \sim 17.2$ mag at that position.}
\end{figure*}

A large database collected during the observing period was planned as:

(a) To search for microlensing events.

(b) To search for variable stars, particularly Cepheids.

(c) To search for other transient events e.g. novae, where a decrease in
brightness of 2 to 5 mag within the first two months of the nova evolution can
easily be detected in our observations.

We have already published a catalog of variable stars which contains data for
26 Cepheids and 333 red variable stars (Joshi et al. 2003a). We are in the
process of analysing some of the possible microlensing candidates found in our
data set (Joshi et al. 2003b, 2003c). Here we report the photometry of two
novae detected in the target field.
\section{Data reduction and photometry}
Standard techniques were used for data reduction using MIDAS and IRAF
software. The dark current correction was not applied due to its
negligible contribution during the maximum exposure time of a frame.
Cosmic rays (CRs) were removed from each frame independently. We added all
the frames of a particular filter taken on a single night and made one frame
per filter per night to increase the signal to noise
ratio. The CCD frames were processed using the DAOPHOT photometry routine
(Stetson 1987). To obtain $R$ and $I$ standard magnitudes, the
photometric calibration was done using Landolt's (1992) standard field
SA98, on a good photometric night of 25/26 October, 2000. A total of 13
secondary stars having $0.09 \le (R-I) \le 1.0$ were observed over a wide range 
of airmasses. The typical error in magnitudes at $\sim$20.0 mag level is
$\sim$ 0.10 and 0.15 mag in $R$ and $I$ bands respectively. In order to use
observations in non-photometric conditions, differential photometry was
done assuming that the errors introduced due to colour difference between nova
and comparison stars were much smaller than the zero point errors. Further
details of the photometric calibration has been given in Joshi et al.
(2003a).

\section{Identification of the novae}
We identified two novae in the target field while searching for
microlensing events using the pixel technique which was initially proposed to
detect microlensing events by Baillon et al. (1993). The implementation of
the pixel technique in our data is described in detail by Joshi et al. (2001,
2003c). The two novae detected in our survey, one in 2000 and the other in 2001
observing seasons, are named nova NMS-1 and nova NMS-2 where NMS is an acronym
for our project `Nainital Microlensing Survey'. The two novae are individually
discussed in the following subsections. 
\subsection {Nova NMS-1}
The nova NMS-1 having celestial coordinates $\alpha _{2000}$ = $00^{h} 42^{m}
57^{s}.1$ and $\delta_{2000}$ = $+41^{\circ}07^{\prime}15^{''}.7$ was reported
in the IAU circular by Donato et al. (2001). When we started observations at
18:54 UT on October 18, 2000, the nova was still brightening. We followed it
through the 2000-2001 observing season. Fig. 1 shows two images where the nova
is unresolved and at maximum brightness.

\setcounter{figure}{1}
\begin{figure}[h]
\includegraphics[height=12.0cm,width=9.0cm]{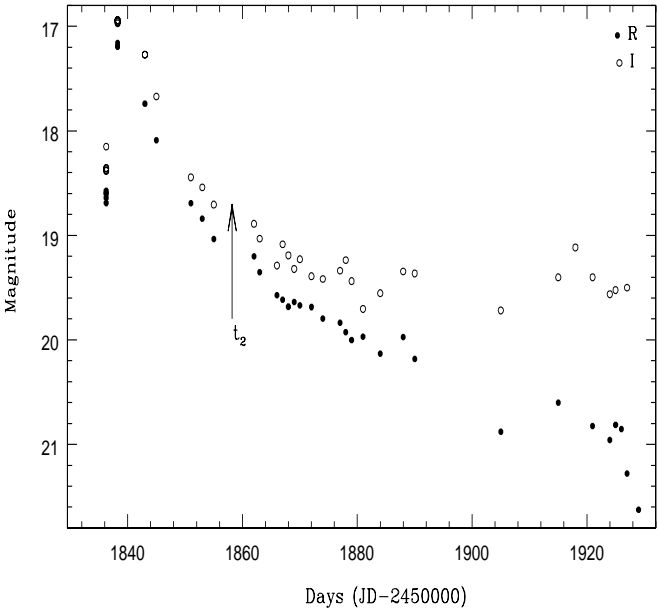}
\vspace{-4.0cm}
\caption{$R, I$ light curves of the nova NMS-1 during outburst where filled and
open circles represent $R$ and $I$ magnitudes respectively. The arrow indicates 
the time where the nova falls 2 mag below its maximum brightness.}
\end{figure}
\setcounter{table}{0}
\begin{table*}
\caption
{\small The $R, I$ magnitudes of the nova NMS-1 during outburst. The fractional value
for the Julian date is given for those points where we calibrated the magnitude
for each individual frame of that night.}
{\scriptsize
\begin{center}
\begin{tabular}{cccc|ccc}\hline
J.D.     &  $R$ &  J.D.    & $I$ &J.D.     &  $R$ &  $I$ \\  
(+2450000)&  (mag)       & (+2450000)&  (mag)  &  (+2450000)&  (mag)       &  (mag)   \\  \hline
1836.287 & 18.69$\pm$0.02 & 1836.316 & 18.38$\pm$0.03&1869     & 19.64$\pm$0.03   & 19.32$\pm$0.06\\
1836.292 & 18.64$\pm$0.02 & 1836.321 & 18.38$\pm$0.02&1870     & 19.67$\pm$0.02   & 19.23$\pm$0.05\\
1836.296 & 18.60$\pm$0.02 & 1836.327 & 18.36$\pm$0.02&1872     & 19.69$\pm$0.03   & 19.39$\pm$0.05\\
1836.299 & 18.57$\pm$0.02 & 1836.333 & 18.36$\pm$0.02&1874     & 19.80$\pm$0.02   & 19.42$\pm$0.05\\
1836.303 & 18.59$\pm$0.02 & 1836.338 & 18.15$\pm$0.05&1877     & 19.84$\pm$0.03   & 19.34$\pm$0.07\\
1838.213 & 17.19$\pm$0.01 & 1838.236 & 16.97$\pm$0.01&1878     & 19.93$\pm$0.02   & 19.24$\pm$0.06\\
1838.218 & 17.16$\pm$0.01 & 1838.245 & 16.95$\pm$0.01&1879     & 20.00$\pm$0.03   & 19.44$\pm$0.08\\
1838.224 & 17.19$\pm$0.01 & 1838.250 & 16.94$\pm$0.01&1881     & 19.97$\pm$0.02   & 19.70$\pm$0.06\\
  -      &    -           & 1838.255 & 16.96$\pm$0.01&1884     & 20.13$\pm$0.02   & 19.55$\pm$0.05\\
  -      &    -           & 1838.259 & 16.94$\pm$0.01&1888     & 19.97$\pm$0.07   & 19.35$\pm$0.08\\
 1843     & 17.74$\pm$0.01&          & 17.27$\pm$0.01&1890     & 20.18$\pm$0.06   & 19.37$\pm$0.05\\
 1845     & 18.09$\pm$0.02&          & 17.67$\pm$0.02&1905     & 20.88$\pm$0.06   & 19.72$\pm$0.05\\
 1851     & 18.69$\pm$0.02&          & 18.44$\pm$0.03&1915     & 20.60$\pm$0.12   & 19.40$\pm$0.12\\
 1853     & 18.84$\pm$0.02&          & 18.54$\pm$0.05&1918     & -                & 19.12$\pm$0.08\\
 1855     & 19.04$\pm$0.02&          & 18.71$\pm$0.04&1921     & 20.82$\pm$0.08   & 19.40$\pm$0.06\\
 1862     & 19.20$\pm$0.03&          & 18.89$\pm$0.04&1924     & 20.96$\pm$0.07   & 19.56$\pm$0.04\\
 1863     & 19.35$\pm$0.03&          & 19.03$\pm$0.04&1925     & 20.81$\pm$0.06   & 19.52$\pm$0.04\\
 1866     & 19.57$\pm$0.02&          & 19.29$\pm$0.07&1926     & 20.85$\pm$0.08   & -             \\
 1867     & 19.61$\pm$0.03&          & 19.09$\pm$0.04&1927     & 21.28$\pm$0.09   & 19.50$\pm$0.07\\
 1868     & 19.68$\pm$0.02&          & 19.19$\pm$0.06&1929     & 21.63$\pm$0.11   & -             \\
 \hline   
\end{tabular}
\end{center}     
}
\end{table*}
The photometry of nova NMS-1 has been carried out using the PSF profile fitting
technique. Since it was sufficiently bright during the early phase of its
eruption and gradually increased its brightness, we carried out photometry on
each
individual frame of the observations of 18 and 20 October, 2000 (see Table 1).
The light curve of nova NMS-1 is shown in Fig. 2. The brightness of the nova
increased by $\sim$ 0.10 mag in the $R$ band from 18:54 UT to 19:17 UT and by
$\sim$ 0.23 mag in $I$ band from 19:35 UT to 20:07 UT. From a comparison of
individual frames taken on that day, it appears that we have captured nova
NMS-1 during its peak brightness. The total brightness of the nova increased by 
$\sim$ 1.5 mag during October 18-20, 2000 followed by an almost exponential
decay. Its rate of decline, $v_d$, defined as the rate in mag day$^{-1}$ at
which a nova drops to two magnitudes below maximum brightness, is estimated to
be:
\begin{equation}
 v_{d,R} \sim 0.11 ~ \tt{mag~day}^{-1}
\end{equation}
\begin{equation}
 v_{d,I} \sim 0.11 ~ \tt{mag~day}^{-1}
\end{equation}
for $R$ and $I$ bands respectively. The observed values of $v_d$ suggest
that the nova \linebreak NMS-1 was a fast nova. The maximum magnitude of the
nova NMS-1
at peak brightness is $M_R$(max) $\sim -7.96$ and $M_I$(max) $\sim -8.02$
mag in $R$ and $I$ bands respectively. Here we consider a true distance
modulus of 24.49 mag for M31 and a total extinction of 0.63 mag and 0.47 mag
towards our observed direction in the $R$ and $I$ bands respectively (cf.
Joshi et al. 2003a). The correlation between observed $R_{max}$ and
rate of decline is consistent with that illustrated by Capaccioli et al.
(1989) for novae in M31.
\setcounter{figure}{2}
\begin{figure*}[ht]
\centering
\includegraphics[height=6.0cm,width=6.0cm]{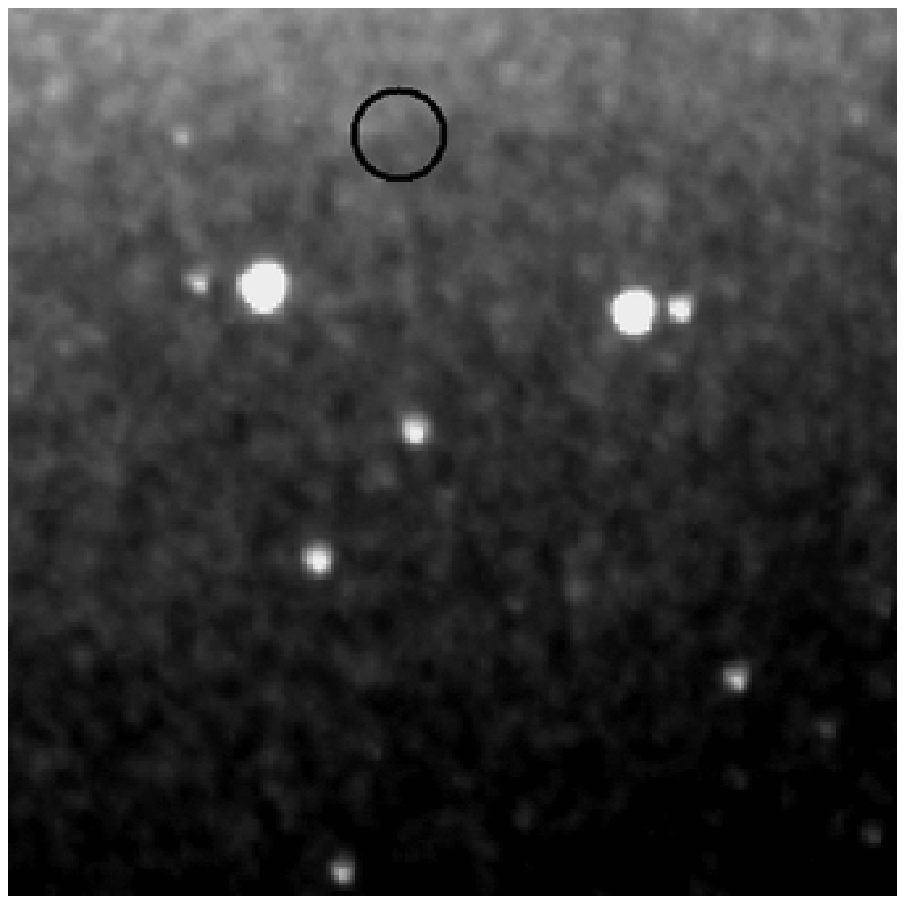} \hspace{1cm} \includegraphics[height=6.0cm,width=6.0cm]{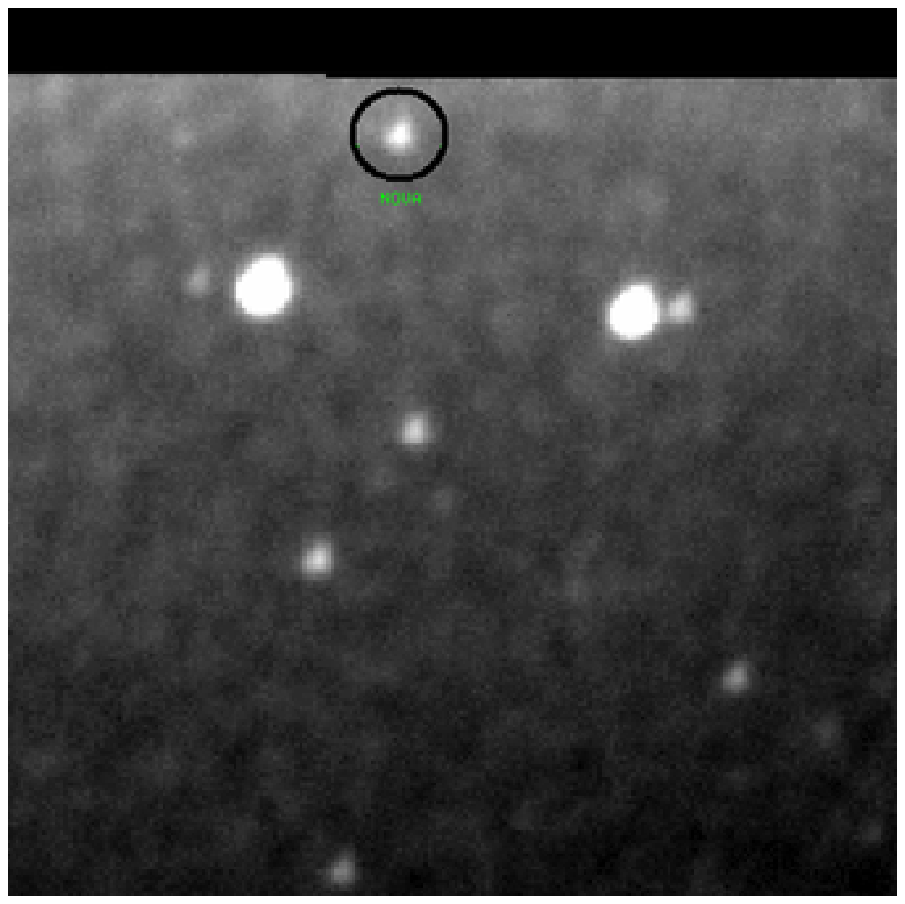}
\caption{A 2 arcmin wide subset of two different $R$ band images taken on
January 17, 2001 (left panel) and October 19, 2001 (right panel). North is at the
top and East is to the left. The left frame shows no star at the position
marked by a circle while right frame shows a star (nova NMS-2) at that position 
of brightness $R \sim$ 17.7 mag.}
\end{figure*}
\subsection {Nova NMS-2}
The nova NMS-2 was already reported in the IAU circular by Li (2001). Fig. 3
shows two images where the nova is at unresolved and maximum brightness phase. The
celestial coordinates of the nova NMS-2 are
$\alpha _{2000}$ = $00^{h} 43^{m} 03^{s}.3$ and
$\delta_{2000}$ = $+41^{\circ}12^{\prime}10^{''}.8$. Since this nova was
situated at the edge of our target field, it could not be observed in all the
images.  When we started our observation on October 12, 2001, it was
already in a descending phase of brightness.

Photometry of the nova was carried out during the brighter phase of
outburst. We give its $R$ and $I$ magnitudes in Table 2 and the light
curve is shown in Fig. 4. A variation of more than 1.5 mag in $R$ band is
seen during the first 35 days of its observations. For the nova NMS-2, our
photometry is insufficient to characterize its speed class since we do not
have any information of its maximum brightness. The brightest magnitude of
the nova NMS-2 in our observations is estimated to be $\sim$ 17.7 mag
in both $R$ and $I$ bands indicating that it must be a star brighter than
$M_R,M_I\sim -7.4$ mag during its peak brightness.

The decay rate for the nova NMS-2 during our observation period is estimated
to be:
\begin{equation}
 \frac {dM(R)}{dt} \sim 0.03\pm0.003 ~ \tt{mag~day}^{-1}
\end{equation}
\begin{equation}
 \frac {dM(I)}{dt} \sim 0.05\pm0.004 ~ \tt{mag~day}^{-1}
\end{equation}
for $R$ and $I$ bands respectively. The small value of the decay rate for
the nova NMS-2 suggests that either it was a slow nova or, most likely, we
observed it very late after it reached its peak brightness.
 
A re-brightening in the light curve of the nova NMS-2 from JD $\sim$ 2452210 to
2452220 is evident in both $R$ and $I$ bands. This type of re-brightening
profile in nova evolution is quite rare but not unique (Bonifacio et al.
2000). However, due to insufficient data points, we are not able to draw
any firm conclusion about its exact behaviour. The evolution of $(R-I)$ colour
for the nova NMS-2 is shown in the lower panel of Fig. 4. It becomes bluer
with time which is consistent with normal nova behaviour. However,
scattering in the data due to poor flat fielding at the edge of frames
prevents us from drawing any conclusion about the dust formation in the ejecta.
\setcounter{figure}{3}
\begin{figure}[h]
\includegraphics[height=11.0cm,width=9.0cm]{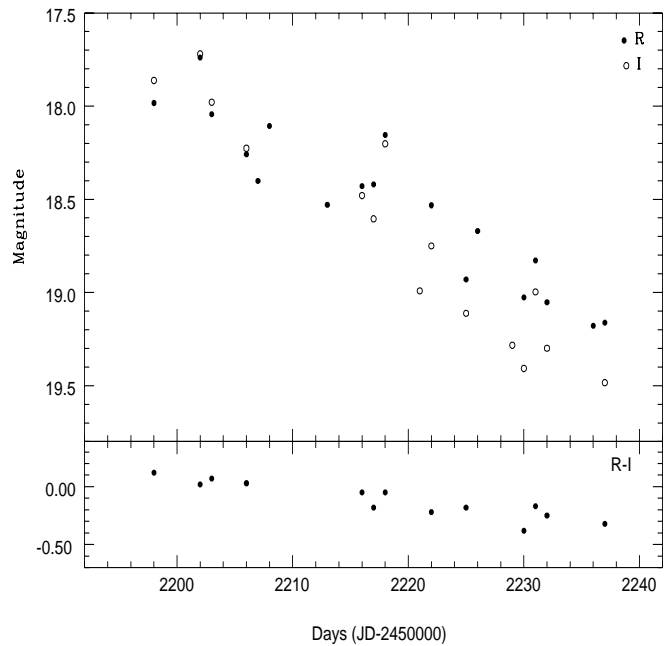}
\vspace{-2.7cm}
\caption{$R, I$ light curve of the nova NMS-2 during outburst is shown in upper panel
where filled and open circles represent $R$ and $I$ magnitudes respectively.
The colour variation of the nova NMS-2 is shown in the lower panel.}
\end{figure}
\setcounter{table}{1}
\begin{table*}
\caption {\small The $R, I$ magnitudes of the nova NMS-2 during outburst.}
{\scriptsize
\begin{center}
\begin{tabular}{ccc|ccc}\hline
J.D.      & $R$  &   $I$  & J.D.     &  $R$  &    $I$\\  
(+2450000)& (mag)&  (mag) &(+2450000)&  (mag)&   (mag)\\  \hline
2198 & 17.98$\pm$0.01 & 17.86$\pm$0.04&2221 &   -            & 18.99$\pm$0.08\\
2202 & 17.74$\pm$0.02 & 17.72$\pm$0.02&2222 & 18.53$\pm$0.03 & 18.75$\pm$0.05\\
2203 & 18.04$\pm$0.02 & 17.98$\pm$0.04&2225 & 18.93$\pm$0.03 & 19.11$\pm$0.06\\
2206 & 18.26$\pm$0.03 & 18.23$\pm$0.03&2226 & 18.67$\pm$0.03 &   -           \\
2207 & 18.40$\pm$0.02 &   -           &2229 &   -            & 19.28$\pm$0.06\\     
2208 & 18.11$\pm$0.02 &   -           &2230 & 19.03$\pm$0.03 & 19.41$\pm$0.07\\     
2213 & 18.53$\pm$0.03 &   -           &2231 & 18.83$\pm$0.04 & 19.00$\pm$0.07\\     
2216 & 18.43$\pm$0.03 & 18.48$\pm$0.05&2232 & 19.05$\pm$0.03 & 19.30$\pm$0.08\\
2217 & 18.42$\pm$0.02 & 18.61$\pm$0.07&2236 & 19.18$\pm$0.03 &   -           \\
2218 & 18.16$\pm$0.03 & 18.20$\pm$0.04&2237 & 19.16$\pm$0.04 & 19.48$\pm$0.15\\
\hline
\end{tabular}
\end{center}
}
\end{table*}

The detection rate of novae in M31 is $\sim$ 30 novae per year (Arp 1956,
Capaccioli et al. 1989, Shafter \& Irby 2001) and 2 novae detected in our
survey, keeping our field of view and total observing time in mind, is
consistent with these estimates.
\section {Discussion}
The $R$ and $I$ magnitudes of the nova NMS-1 at the brightest phase are
estimated to be -7.96 and -8.02 mag respectively with a rate of decline of
$\sim$ 0.11 mag day$^{-1}$. A correlation of the rate of decline with the
observed peak flux is in good agreement with the maximum magnitude versus
rate of decline (MMRD) curve given by Capaccioli et al. (1989) for M31 novae. 

It is instructive to analyse the light curves and colour index, to investigate
the evolution of the remnant. From the observed light curve of the nova NMS-1
(Fig. 2), it is evident that the flux in $R$ and $I$ bands decline in a similar
way for the first three weeks but the late time evolution is qualitatively
different. A difference in the profile of the light curves in two bands is a
valuable diagnostic of the physical processes operating in the ejecta. We have
neither spectroscopic information nor infra-red photometry, nevertheless, we
might estimate the possible evolution of the expanding shell from the $R, I$
light curves.

To study the photometric behavior of the nova NMS-1, we divided the declining
part of the light curve into three sections as shown in Fig. 5(a1). The initial
rapid decline in the flux follows a similar pattern in $R$ and $I$ bands and
the colour
remains practically unchanged; essentially this is the phase where emission
from a region of progressively lower effective radius is received. However,
after about three weeks, the ($R-I$) colour shows a gradual increase indicating
that the region is cooling. After about 40 days, the $R$ band flux is still
progressively declining similar to that observed for the nova Herculis
1991 (Harrison \& Stringfellow 1994) in the $V$ band. The $I$ band, however,
probably shows a slight gradual increase in flux during this period; 
a rapid rise in the colour index of the nova is seen in Fig. 5(a2).
A similar behaviour has been observed in the case of nova Herculis 1991 where
an almost constant plateau is seen in $J$, $H$, $K$ and $L$ bands (see Fig.
5(b1) and 5(b2)). Although, we have neither sufficient observations in the
later stage nor supporting spectroscopy, we still believe that probably there
is a signature of the formation of neutral hydrogen. The decrease in $R$ band
flux but the reverse
trend in $I$ band flux is likely to be due to the $H_\alpha$ absorption and
re-radiation in longer wavelengths. There is some reason to speculate this
because: a) we see a nearly constant peak Balmer flux of $10^{37}$ ergs/sec
declining rapidly, b) moderately large velocity gradient in the shell is shown
by Balmer lines and c) A Balmer jump in the spectra (cf. Downes et al. 2001)
are seen. If such a scenario can be confirmed, it could provide a diagnostic of
dust formation during the expansion of wind in the nova.
\setcounter{figure}{4}
\begin{figure*}[ht]
\centering
\includegraphics[height=11.0cm,width=14.0cm]{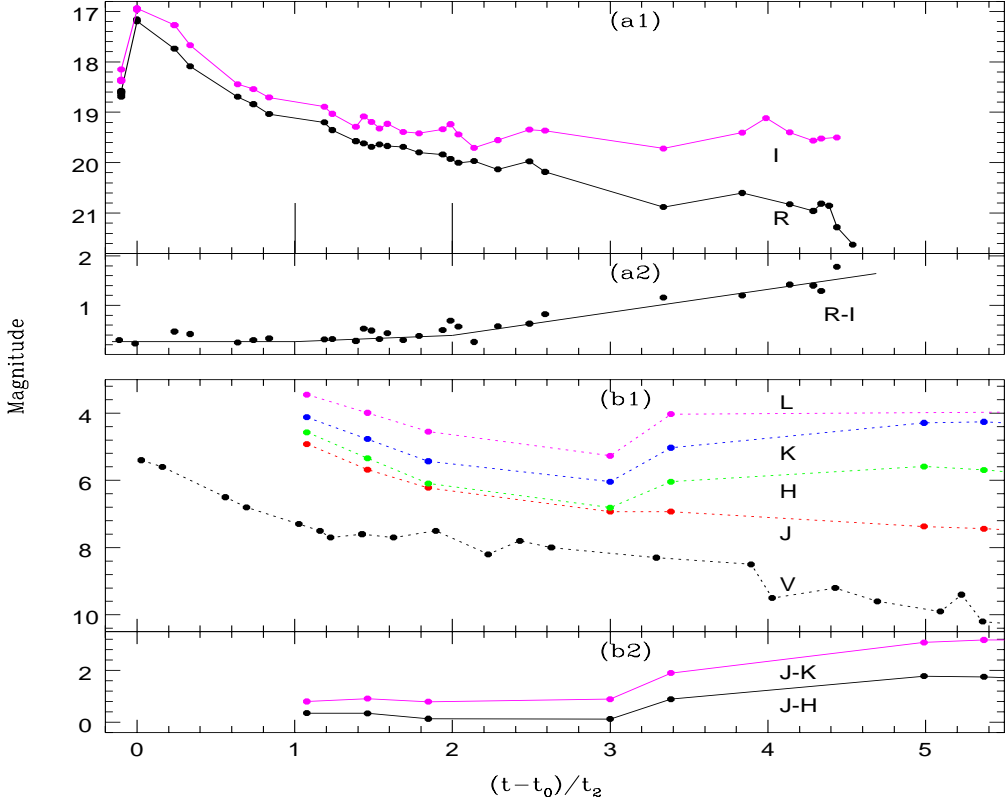}
\caption{$R, I$ light curves of the nova NMS-1 and its colour variation during the
outburst is shown in the upper panels (a1) and (a2) respectively. $VJHKL$ data 
along with the colour variation of the nova Herculis 1991, taken from
the literature, is plotted in the lower panels (b1) and (b2) respectively, to
compare the behaviour of the two novae. Here $t$, $t_0$ and $t_2$ indicate the
time of observation, time of the maximum brightness and rate of decline
respectively.}
\end{figure*}

For the nova NMS-2, we do not have light curve measurements in the initial
phase and hence
the epoch of peak brightness cannot be estimated. Unlike the nova NMS-1 where
we had observations for about three months, the nova NMS-2 could be monitored 
only for about 40 days and during this period, $R$ and $I$ flux 
show a similar trend, except that $I$ declines faster. This is consistent with
emission coming from a region of lower radius but higher temperature. Probably, 
the flux in both bands shows a bump similar to that observed in some novae.

What does our light curves imply about the formation of dust in the nova wind? 
Novae are believed to be a source of dust enrichment in the interstellar
medium. However, there are two conflicting views (cf. Bode \& Evans, chap9):
(1) Graphite or silicate grains can be formed in the cooling wind of nova
depending on the chemical composition of white dwarf.
(2) The nova environment could be hot and might not be the conducive of dust
formation, in which case dust grains already present in the shell could cause
an infra-red excess. If the expanding cool wind in the nova can produce grains,
we should be able to see the reduction in temperature followed by an infra-red
excess. A good monitoring program that tracks the temperature, luminosity as
well as possible spectroscopy will be helpful in settling the question. Our
limited $R, I$ observations suggest that the possible neutral hydrogen
formation, when ($R-I$) increases to about 1 mag from near zero, could act
to shield dust grains against evaporation. This might be the reason for
the slow increase in $I$ band flux despite the drop in $R$ band flux
which is contributed in part by regions bluewards of the Balmer alpha line.
\section {Conclusion}
We found two novae in our survey carried out during 1998 to 2001 for an
$\sim 13'\times13'$ field belonging to the M31 disk population. The
rate of decline of nova NMS-1 suggests that it was a fast nova while
the decay rate for nova NMS-2 suggests that either it was a slow nova or
we observed it at a very late stage after the maximum brightness phase. The
nova NMS-1 becomes cooler with time about six weeks after the maximum
brightness phase accompanied by a slow increase in $I$ band flux, although
the $R$ band flux
still continues to decrease. This is in contrast with the normal
behaviour of the novae. We suspect a secondary bump in $R, I$ light
curves of the nova NMS-2, which is not unusual for normal nova evolution.

~

{\it Acknowledgments}
We are grateful to St\'{e}phane Paulin-Henriksson for his help. We thank
the anonymous referee for the useful comments. This study is a part of the
project 2404-3 supported by Indo-French center for the Promotion of Advanced
Research, New Delhi.


\begin{thebibliography}{}
\bibitem{} Arp, H., 1956, AJ, 61, 15
\bibitem{} Baillon, P., Bouquet, A., Kaplan J. \& Giraud-H\'{e}raud Y., 1993,
A\&A, 277, 1.
\bibitem{} Bode, M.F. \& Evans, A., 1989, Classical Novae, Chap. 9, (Publisher, John Willy \& Sons Ltd)
\bibitem{} Bonifacio, P., Selvelli, P.L. \& Caffau, E., 2000, A\&A, 356, L53
\bibitem{} Ciardullo, R., Ford, H.C., Neil, J.D., Jacoby, G.H. \& Shafter, A.W., 1987,
ApJ, 318, 520
\bibitem{} Capaccioli, M., Della Valle, M., D'Onofrio, M. \& Rosino, L., 1989, AJ,
97, 1622
\bibitem{} Donato, L., Garzvia, S., Conano, V., Sostero, G. \& Korlevic, K., 2001, IAUC 7516
\bibitem{} Downes, R.A., Duerbeck, H.W. \& Delahodde, C.E., 2001, JAD, 7, 6
\bibitem{} Harrison, T.E. \& Stringfellow, G.S., 1994, ApJ, 437, 827
\bibitem{} Hubble, E., 1929, ApJ, 69, 103
\bibitem{} Jacoby, C.H., Branch, D.G., Ciardullo, R., et al., 1992, PASP, 104, 599
\bibitem{} Joshi, Y.C., Pandey, A.K., Narasimha, D. \& Sagar, R., 2001, BASI, 29, 531
\bibitem{} Joshi, Y.C., Pandey, A.K., Narasimha, D., Sagar, R. \& Giraud-H\'{e}raud,
Y., 2003a, A\&A, 402, 113
\bibitem{} Joshi, Y.C., Pandey, A.K., Narasimha, D. \& Sagar, R., 2003b, BASI,
presented in XXII ASI Meeting, Feb. 13-15, 2003
\bibitem{} Joshi, Y.C., et al., 2003c, under preparation
\bibitem{} Landolt, A.U., 1992, AJ, 104, 340
\bibitem{} Li, W.D., 2001, IAUC 7729
\bibitem{} McLaughlin, D.B., 1945, PASP, 57, 69
\bibitem{} Rector, T.A., Jacoby, G.H., Corbett, D.L., Denham, M., RBSE Nova Search Team
1999, BAAS, 195, 36.08
\bibitem{} Rosino, L., 1964, A\&A, 27, 498
\bibitem{} Rosino, L., 1973, A\&ASS, 3, 347
\bibitem{} Rosino, L., Capacciol, M., D'Onofrio, M., \& Della Valle, M., 1989, AJ, 97,
83
\bibitem{} Sharov, A.S. \& Alksnis, A., 1991, ApSS, 180, 273
\bibitem{} Shafter, A.W. \& Irby, B.K., 2001, ApJ, 563, 749
\bibitem{} Stetson, P.B., 1987, PASP, 99, 191
\bibitem{} Tomaney, A.B. \& Shafter, A.W., 1992, ApJS, 81, 683
\bibitem{} Van den Bergh, S., 1988, PASP, 100, 1486
\end{thebibliography}
\end{document}